\documentclass[seceq,preprint]{ptptex}

\newcommand{\dz}{\frac{dz}{2\pi i}}
\newcommand\pp{\pi^+}
\newcommand\pb{{\bar\pi}}
\newcommand\ld{\lambda^2}
\newcommand\lb{{\bar\lambda}^{\dot 2}}
\newcommand\dth{\partial\theta}
\newcommand\dthb{\partial{\bar\theta}}
\newcommand\Sb{\bar{S}}
\newcommand\dw{\oint\frac{dw}{2\pi i}}
\newcommand{\e}{\epsilon^{ij}}
\notypesetlogo                       
\preprintnumber[3cm]{
YITP-07-38\\
KEK-TH-1161
\\ June, 2007}

\title{
Lorentz Anomaly
in Semi-Light-Cone Gauge Superstrings
}

\author{
Hiroshi \textsc{Kunitomo}$^1$
and
Shun'ya \textsc{Mizoguchi}$^{2,3}$%
}

\inst{
$^1$Yukawa Institute for Theoretical Physics,
Kyoto University,\\ Kyoto 606-8502, Japan\\%
$^2$High Energy Accelerator Research Organization (KEK),\\
Tsukuba 305-0801, Japan\\
$^3$Department of Particle and Nuclear Physics,\\
The Graduate University for Advanced Studies,\\
Tsukuba 305-0801, Japan
}

\abst{
We study the Lorentz invariance of $D=4$ and 6 superstrings in the double-spinor 
formalism, which are equivalent to the $D=4$ and 6 superstrings 
in the pure-spinor formalism in the sense of the BRST cohomology.
We first re-examine how the conformal and Lorentz anomalies appear in the $D=4$ 
and 6 Green-Schwarz superstrings in the semi-light-cone gauge in the framework of 
BRST quantization. We construct a set of BRST invariant Lorentz generators 
and show that they do not form a closed algebra, even cohomologically. We then turn 
to the construction of Lorentz generators in the $D=4$ and 6 double-spinor 
superstrings, and show that the Lorentz invariance is again anomalous. We also discuss 
the relation between the anomaly-free Lorentz generators in the lower-dimensional 
pure-spinor formalisms and that obtained in this paper. 
}

\begin{document}
\maketitle

\section{Introduction}
Recently, it has been recognized that the covariant quantization of superstrings
using pure spinors \cite{B} can be naturally understood in terms of a
Green-Schwarz-like superstring with twice as many fermionic degrees of freedom, 
the double-spinor (DS) formalism \cite{AK}. The superstring in the DS formalism 
possesses an additional local symmetry, and is classically gauge equivalent to the 
ordinary Green-Schwarz (GS) superstring. 
Imposing the semi-light-cone gauge condition on one half of the fermionic variables,  
Aisaka and Kazama completed a Dirac/BRST 
quantization of the $D=10$ DS superstring, finding that the resulting system is 
cohomologically equivalent to the PS superstring.\footnote{
See \citen{Mazzucato} for a different formulation which also relates
GS and PS superstrings.} 
In this way, they uncovered the ``origin" of  the formalism, and, in particular, 
they derived the previously mysterious seventeen first-class constraints\cite{BM}
assumed to clarify the relation between GS and PS superstrings.

In a previous paper, Ref.~\citen{KM}, we applied this idea to lower-dimensional 
($D=4$ and 6) cases \cite{GW,W}. The primary motivation of that work was to understand
how the concept of the critical dimension emerges in the PS formalism. We 
have shown that, starting from similar Lagrangians, 
$D=4$ and $D=6$ DS superstrings can be 
BRST quantized to yield free CFTs similar to 
the semi-light-cone gauge GS superstrings, along with 
additional conjugate pair systems and extra constraints.
The BRST charges again reduce
to those of the lower-dimensional
PS superstrings through similarity transformations.

Thus, the DS superstrings ``interpolate" between the GS and PS 
superstrings, but this raises some questions.
The GS superstring theories  
have a Lorentz anomaly in lower dimensional cases,
while the PS superstring theories have anomaly-free Lorentz 
generators\cite{GW,W}. 
Where does this difference come from?
Then, as a related  
question, what do ``quantum mechanically consistent 
$D=4$ and 6 superstrings"  describe? 

The DS superstrings are closely related to the GS 
superstrings in the semi-light-cone gauge\cite{AK}.
The presence or absence of Lorentz and conformal anomalies for the $D=10$ semi-light-cone 
gauge GS superstring was a subject of great debate
in the late 1980s and early 1990s. 
In Ref.~\citen{KR}, it was revealed that, contrary to the prevailing
belief at that time\cite{earlierbelief}, 
the $D=10$ GS superstring in the semi-light-cone 
gauge has a non-vanishing conformal anomaly.  
Later, it was shown that this conformal anomaly is canceled 
by introducing a certain local counterterm, 
and the Lorentz algebras become closed with a suitable modification of
the Lorentz generators\cite{BvNP,PvN}.
This local counterterm can be viewed as a coupling to a certain dilaton background.
More recently, the Lorentz invariance 
of the $D=10$ GS superstring in the semi-light-cone gauge 
has been re-examined and proved using the BRST method \cite{BM}. 

In this paper, 
we first examine the conformal and Lorentz anomalies of the $D=4$ and 6 GS 
superstrings in the semi-light-cone gauge. 
We  BRST quantize these lower-dimensional GS superstrings
in a manner similar to that for the DS superstring in
Ref.~\citen{KM}. The key step
in this procedure is 
the modification of the quantum constraints, and we argue that it effectively 
changes the background from a flat space-time to a linear-dilaton-like one. 
We then construct a set of BRST invariant Lorentz generators and show 
that they are not closed, as expected.

Next, we turn to an examination of the Lorentz invariance of the $D=4$ and 6 DS superstrings 
studied in Ref.~\citen{KM}. We present a complete set of BRST-invariant Lorentz generators 
in both cases. We then show that they form the correct Lorentz algebra,
except for the commutators between the ``$i-$" generators, which, again, are not BRST exact. 
Finally, we investigate the relation between these charges and the anomaly-free Lorentz 
generators in the $D=4$ PS formalism described in Refs.~\citen{GW} and \citen{W}. 

The organization of this paper is as follows.  
In \S 2, we study the conformal and Lorentz anomalies of the $D=4$ and 6 
GS superstrings 
in the semi-light-cone gauge using the BRST method. 
We derive the BRST-invariant Lorentz generators of the semi-light-cone
gauge DS superstring and compute their algebras in \S 3. 
In the final section, we discuss the difference between the anomaly-free Lorentz 
generators of  Refs. \citen{GW} and \citen{W} and those obtained in this paper.

\section{The Lorentz invariance of lower-dimensional GS superstrings in the semi-light-cone gauge}
\subsection{The $D=4$ GS superstring in the semi-light-cone gauge}

The Lagrangian of the $D=4$ Green-Schwarz (GS) superstring is 
an obvious generalization of the $D=10$ GS Lagrangian,\cite{AK} with 
an appropriate spinor structure in four dimensions:
\begin{subequations}
\begin{eqnarray}
\mathcal{L}&=&\mathcal{L}_K+\mathcal{L}_{WZ},\\
\mathcal{L}_K&=&
-\frac{1}{2}\sqrt{-g}g^{ab}\Pi^\mu_a\Pi_{\mu b},\\
\mathcal{L}_{WZ}&=&
\epsilon^{ab}\Pi^\mu_a(W_{\mu b}-\hat{W}_{\mu b})
-\epsilon^{ab}W^\mu_a\hat{W}_{\mu b}
\end{eqnarray}
\end{subequations}
with
\begin{eqnarray}\label{Pi_aGS}
 \Pi^\mu_a&=&\partial_aX^\mu
-\sum_{A=1}^2W^{A\mu}_a,\\
 W^{A\mu}_a&=&i\theta^A\sigma^\mu\partial_a\bar{\theta}^A
-i\partial_a\theta^A\sigma^\mu\bar{\theta}^A.\label{WAGS}
\end{eqnarray}
Here, we employ the notation used in Ref.~\citen{KM}:
$\mu,\nu=0,1,2,3$ are the flat space-time 
indices with the metric $\eta^{\mu\nu}=\mbox{diag}[+1,-1,-1,-1]$,
$a,b=0,1$ are the worldsheet indices,
$\sigma^\mu$ are the two-by-two hermitian off-diagonal 
blocks of the gamma matrices in the chiral representation, 
$\theta^A$ are complex Weyl spinors, with 
$A=1,2$ labeling the left and right degrees of freedom 
after the semi-light-cone 
gauge fixing. We also adopt the notation $W^{\mu}_a= W^{A=1,\mu}_a$, 
$\hat{W}^{\mu}_a= W^{A=2,\mu}_a$, etc. 

The fermionic constraints are simply
\begin{subequations}\label{primary4GS}
\begin{eqnarray}
D^A_\alpha&=&k^A_\alpha
-i(k^\mu+\eta_A(\Pi^\mu_1+W^{\mu\bar{A}}_1))
(\sigma_\mu\bar{\theta}^A)_\alpha\approx0, \label{DA}\\
\bar{D}^A_\alpha&=&\bar{k}^A_{\dot{\alpha}}
-i(k^\mu+\eta_A(\Pi^\mu_1+W^{\mu\bar{A}}_1))
(\theta^A\sigma_\mu)_{\dot{\alpha}}\approx0,
\end{eqnarray}
\end{subequations}
where $\bar A=1(2)$ if $A=2(1)$. 
Parameterizing the worldsheet metric as 
\begin{eqnarray}
g_{ab}&=&\left(
\begin{array}{cc}
-N^2+\gamma (N^1)^2&\gamma N^1\\
\gamma N^1&\gamma
\end{array}
\right),
\end{eqnarray}
in the ADM form, we obtain the Hamiltonian 
\begin{eqnarray}
\mathcal{H}&=&
\frac{N}{\sqrt{\gamma}}T_0+N^1T_1
+\dot{\theta}^{A\alpha}D^A_\alpha
+\dot{\bar{\theta}}^{A\dot{\alpha}}\bar{D}^A_{\dot{\alpha}},
\label{hamiltonian4}
\end{eqnarray}
where 
\begin{subequations}\label{T+-GS} 
\begin{eqnarray}
 T_+&=&\frac{1}{2}(T_0+T_1)=\frac{1}{4}\Pi^\mu\Pi_\mu, \label{T+GS}\\
 T_-&=&\frac{1}{2}(T_0-T_1)=\frac{1}{4}\hat{\Pi}^\mu\hat{\Pi}_\mu,
\end{eqnarray}
\end{subequations}
\begin{subequations} 
\begin{eqnarray}
 \Pi^\mu
&=&k^\mu+X'^\mu
-2W^\mu_1,\label{Pi^muGS}
\\
\hat{\Pi}^\mu
&=&k^\mu-X'^\mu
+2\hat{W}^\mu_1. 
\end{eqnarray}
\end{subequations}
In fact, all the above formulas can be derived from the corresponding 
ones in the $D=4$ DS formalism \cite{KM} 
by setting all the variables with tildes to zero.
Assuming the Poisson brackets  
\begin{subequations}\label{Poisson4}
\begin{eqnarray}
\{X^\mu(\sigma),k^\nu(\sigma')\}_P&=&
\eta^{\mu\nu}\delta(\sigma-\sigma'),\label{Poisson4Xk}\\ 
\{\theta^{A\alpha}(\sigma),k^B_\beta(\sigma')\}_P&=&
-\delta^{AB}\delta^\alpha_\beta\delta(\sigma-\sigma'),\label{Poisson4thetak}\\ 
\{\bar{\theta}^{A\dot{\alpha}}(\sigma),\bar{k}^B_{\dot{\beta}}(\sigma')\}_P&=&
-\delta^{AB}\delta^{\dot{\alpha}}_{\dot{\beta}}\delta(\sigma-\sigma'),\label{Poisson4thetabark}
\end{eqnarray}
\end{subequations}
we find that two of the four fermionic 
constraints are first class, generating the kappa symmetry,
and the other two are second class.
Imposing the semi-light-cone gauge condition 
\begin{equation}
{\theta}^2\approx{\bar{\theta}}^{\dot{2}}\approx0,
\label{slcgaugeGS}
\end{equation}
the kappa symmetry is fixed, and all the fermionic constraints become 
second class. Then, the only first-class constraints are the left and right Virasoro 
constraints generated by (\ref{T+-GS}). 

The Dirac bracket can be computed straightforwardly, and the result is 
identical to the DS superstring given in Ref.~\citen{KM}, with 
the variables with tildes replaced by variables without tildes, 
and $T$ and $\Pi^+(\equiv\Pi^0+\Pi^3)$ replaced by variables 
appropriate for the GS superstring.
In this case, unlike in the case of the DS superstrings, 
the Dirac brackets among $X^\mu$ and $k^\nu$
remain canonical; the only necessary modifications  are the familiar 
rescalings 
\begin{equation}
S\equiv\sqrt{2\Pi^+}\theta^1,\qquad
\bar{S}\equiv\sqrt{2\Pi^+}{\bar{\theta}}^{\dot1},
\end{equation}
which satisfy the relations 
\begin{subequations}\label{XSSbarDirac}
\begin{eqnarray}
\{S(\sigma),\bar{S}(\sigma')\}_{D}&=&i\delta(\sigma-\sigma'),\\ 
\{X^\mu(\sigma),S(\sigma')\}_{D}&=&0,\\
\{X^\mu(\sigma),\bar{S}(\sigma')\}_{D}&=&0.
\end{eqnarray}
\end{subequations}

We now turn to the quantization of the $D=4$ GS superstring. 
As in Ref.~\citen{KM},
we replace the Dirac brackets obtained above with appropriate OPEs. 
With some rescalings, the left constraint, $T_0+T_1$, 
becomes the energy-momentum 
tensor $T_{\rm matter}(z)$ composed of free fields:
\begin{eqnarray}
T(z)&=&\frac12\partial X^\mu \partial X_\mu  
- \frac12 (S\partial \bar S - \partial S \bar S).
\end{eqnarray}
The OPEs for the basic holomorphic fields are
\begin{subequations}
\begin{eqnarray}
 X^\mu(z)X^\nu(w)&\sim&\eta^{\mu\nu}\log(z-w),\label{XX}\\
S(z)\bar{S}(w)&\sim&\frac{1}{z-w}.\label{SSbar}
\end{eqnarray}
\end{subequations}
Because the central charge of $T(z)$ is 5,
the ghost contribution $-26$ cannot be cancelled
in four dimensions.
To compensate for the shortage,
we modify the energy-momentum tensor $T(z)$ 
similarly to that in Ref.~\citen{KM}, as
\begin{eqnarray}\label{Tmodified}
T(z) \rightarrow \check{T}(z)&=&\frac12
\partial X^\mu \partial X_\mu 
- \frac12(S\partial \bar S -\partial S \bar S)
+\frac78\partial^2\log \partial X^+,
\end{eqnarray}
with $\eta^{+-}=2$, $\eta^{ij}=-\delta^{ij}$.
In general, a family of energy-momentum tensors 
\begin{eqnarray}\label{TX+X-}
T_{X^+X^-}(z)&=&\frac12\partial X^+ \partial X^- + \xi~\partial^2 \log \partial X^+
\end{eqnarray}
with a parameter $\xi$ has central charge
\begin{eqnarray}
c(\xi)&=&1+24\xi
\end{eqnarray}
if $X^+(z)X^-(w)\sim+2\log(z-w)$. 
Therefore, the logarithm term correctly shifts the 
central charge to 26.
Using this modified energy-momentum tensor, we can construct a standard 
nilpotent BRST charge:
\begin{eqnarray}
Q_{\rm GS}&=&\oint\frac {dz}{2\pi i}\left(c\check{T}+
bc\partial c
\right).
\end{eqnarray} 

Note that although this modification of the energy-momentum tensor may
seem \textit{ad hoc},
it {\em is} required even in the $D=10$ GS superstring in the semi-light-cone gauge. 
Indeed, a one-loop 
analysis reveals the existence of a conformal anomaly of $c=-12$, including the 
$bc$ ghosts, which can only be canceled with a special dilaton coupling 
introduced as a local counterterm \cite{BvNP,PvN}. This causes
a change of the energy-momentum tensor as in (\ref{Tmodified}), though with a 
coefficient of $1/2$ instead of $7/8$. The inclusion of the counterterm also results 
in a modification of the spacetime Lorentz transformation rules, which have been 
shown to have no anomaly.\cite{BvNP,PvN}$^,$\cite{BM} 
Similarly, we can add a  local counterterm to the $D=4$ GS action 
so that the total conformal anomaly vanishes, and this gives rise to a change of 
the energy-momentum tensor (\ref{Tmodified}). 
The question is whether, with that counterterm, the rigid Lorentz symmetry is 
preserved in the theory. Below we examine this point.

A Lorentz generator for the GS superstrings 
in the semi-light-cone gauge basically consists of a Noether current
and, if it does not preserve the 
semi-light-cone gauge condition (\ref{slcgaugeGS}), an additional, compensating 
kappa-symmetry current. In addition, we need some extra terms for the 
BRST invariance of the generators. For the $D=4$ case, we find
\begin{subequations}
\begin{eqnarray}
N^{ij}&=&\frac14\left(-X^i\partial X^j +X^j\partial X^i +i\epsilon^{ij}S\bar S\right)
,\\
N^{+-}&=&\frac14\left(-X^+\partial X^- +X^-\partial X^+ \right)
,\\
N^{i+}&=&\frac14\left(-X^i\partial X^+ +X^+\partial X^i \right)
,\\
N^{i-}&=&\frac14\left(-X^i\partial X^- +X^-\partial X^i+2i\epsilon^{ij}\frac{\partial X^j}{\partial X^+}S\bar S
-\frac72 \frac{\partial^2 X^i}{\partial X^+}\right),\label{Ni-GS}
\end{eqnarray}
\end{subequations}
where $i,j=1,2$ and $\epsilon^{12}=-\epsilon^{21}=1$, $\epsilon^{11}=\epsilon^{22}=0$.
The third term in $N^{i-}$ (\ref{Ni-GS}) comes from the compensating kappa 
transformation, and the fourth term is required for the BRST 
invariance.\footnote{An analogous term is also needed 
for the $D=10$ GS superstring. In this case, one must add
$+\frac{\partial^2 X^i}{\partial X^+}$ to $N^{i-}$ in Eq.~(3.6) of 
Ref.~\citen{BM}.
} 
Lorentz generators constructed from these currents all commute with $Q_{\rm GS}$. 
Defining the charges as
\begin{eqnarray}
M^{\mu\nu}&=&\oint\frac{dz}{2\pi i}N^{\mu\nu}(z),
\end{eqnarray}
it can be verified that they form the $D=4$ Lorentz algebra, except for 
${[}M^{1-},~M^{2-}{]}$,
which is given by
\begin{eqnarray}
{[}M^{1-},~M^{2-}{]}&=&\oint \frac{dz}{2\pi i}\left(
i\frac{S\bar S}{(\partial X^+)^2}	\left(
	\frac12\partial X^\mu \partial X_\mu -\frac78 \frac{\partial^3 X^+}{\partial X^+}
	+\frac74 \frac{(\partial^2 X^+)^2}{(\partial X^+)^2}
	\right)\right.\nonumber\\
&&\left.\hspace{8ex}	-
	\frac34\frac{\partial X^1\partial^2 X^2-\partial^2X^1 \partial X^2}{(\partial X^+)^2}
\right).
\label{M1-M2-}
\end{eqnarray}
Unlike the $D=10$ GS superstring analyzed in Ref.~\citen{BM},
the right hand side cannot be BRST-exact. 
This can be proven as follows.
Suppose that the terms proportional to $S\bar S$ in (\ref{M1-M2-})
could be written as a commutator of $Q_{\rm GS}$ and some BRST ``parent." 
Then, since $\check T$ does not have such a term, 
the parent itself must contain $S\bar S$. 
It is not difficult to show that the only possible choice is  
$\frac{bS\bar S}{(\partial X^+)^2}$ multiplied by some constant. 
However, we have 
\begin{eqnarray}
\left[Q_{\rm GS},~\oint\frac{dz}{2\pi i}i\frac{bS\bar S}{(\partial X^+)^2}(z)\right]
&=&\oint \frac{dz}{2\pi i}\left(
i\frac{S\bar S}{(\partial X^+)^2}	\left(
	\frac12\partial X^\mu \partial X_\mu -\frac18 \frac{\partial^3 X^+}{\partial X^+}
	-\frac78 \frac{(\partial^2 X^+)^2}{(\partial X^+)^2}
	\right)\right.\nonumber\\
&&\left.\hspace{8ex}	+
	\frac34 i\frac{S\partial^2 \bar S +\partial^2 S \bar S}{(\partial X^+)^2}
\right),
\end{eqnarray}
which is inconsistent. Thus, we have shown that (\ref{M1-M2-}) does not vanish, 
even cohomologically, and therefore the Lorentz invariance is broken.
This is a natural result,
because we know that the Lorentz algebra is not closed in the light-cone 
quantization, and this should be independent of the gauge choice.

\subsection{The $D=6$ GS superstring in the semi-light-cone gauge}
The BRST quantization of the $D=6$ GS superstring in the semi-light-cone gauge 
is completely analogous, and therefore we give only a brief summary.
Again, the Dirac brackets for the $D=6$ GS superstring are 
derived from the $D=6$ DS superstring \cite{KM} by similar replacements. 
The matter energy-momentum tensor is 
given by
\begin{eqnarray}
T(z)&=&\frac12
\partial X^\mu \partial X^\mu  
- \frac12 S_a^I\partial S_I^a.
\end{eqnarray}
The relevant OPEs are
\begin{subequations}
\begin{eqnarray}
 X^\mu(z)X^\nu(w)&\sim&\eta^{\mu\nu}\log(z-w),\label{XX6D}\\
S_I^a(z) S_J^b(w)&\sim&-\frac{\epsilon_{IJ}\epsilon^{ab}}{z-w}.\label{SS6D}
\end{eqnarray}
\end{subequations}
Again, we modify the energy-momentum tensor to
\begin{eqnarray}
\check{T}(z)&=&\frac12
\partial X^\mu \partial X^\mu  
-\frac12 S_a^I\partial S_I^a
+\frac34\partial^2\log\partial X^+,
\end{eqnarray}
so that the BRST charge 
\begin{eqnarray}
Q_{\rm GS}&=&\oint\frac {dz}{2\pi i}\left(c\check{T}+
bc\partial c
\right)
\end{eqnarray} 
becomes nilpotent.
The BRST-invariant Lorentz generators are found to be
\begin{subequations}
\begin{eqnarray}
N^{ij}&=&\frac14\left(-X^i\partial X^j +X^j\partial X^i +\frac i2 (S^I \gamma^{ij} S_I)\right)
,\\
N^{+-}&=&\frac14\left(-X^+\partial X^- +X^-\partial X^+ \right)
,\\
N^{i+}&=&\frac14\left(-X^i\partial X^+ +X^+\partial X^i \right)
,\\
N^{i-}&=&\frac14\left(-X^i\partial X^- +X^-\partial X^i+i\frac{\partial X^j}{\partial X^+}(S^I\gamma^{ij} S_I)
-3 \frac{\partial^2 X^i}{\partial X^+}\right).\label{Ni-6dGS}
\end{eqnarray}
\end{subequations}
It can be verified that they form the correct $D=6$ Lorentz algebra, except that 
\begin{eqnarray}
&&{[}M^{i-},~M^{j-}{]}
\nonumber\\
&=&\oint \frac{dw}{2\pi i}\left(
\frac i2 \frac{(S^I \gamma^{ij} S_I)}{(\partial X^+)^2}\left(
\frac12\partial X^\mu \partial X_\mu
-\frac18 S_b^J\partial S_J^b
-\frac34
\frac{\partial^3 X^+}{\partial X^+}
+ 
\frac{(\partial^2 X^+)^2}{(\partial X^+)^2} 
\right)\right.\nonumber\\
&&\hspace{8ex}\left.
-\frac12
\frac{\partial X^i\partial^2 X^j-\partial^2 X^i\partial X^j}{(\partial X^+)^2}
+\frac i4 \frac{(S^I \gamma^{ij} \partial^2 S_I)}{(\partial X^+)^2}
\right.\nonumber\\
&&\hspace{8ex}\left.
+\frac i8 \frac{(S^I\gamma^{ij}\partial S_J)}{(\partial X^+)^2} S_{b}^JS_I^b 
\right).
\label{Mi-Mj-6dGS}
\end{eqnarray}
Again, the right hand side is not BRST-exact: 
As in the $D=4$ case,
the $S$-bilinear terms can only arise from 
a product of $ c\check{T}$ and something proportional to 
$\frac{S^I\gamma^{ij}S_I}{(\partial X^+)^2}$, but we have
\begin{eqnarray}
&&\left[Q_{\rm GS},~\oint\frac{dz}{2\pi i}
i\frac{b(S^I\gamma^{ij} S_I)}{2(\partial X^+)^2}(z) \right]
\nonumber\\
&=&
\oint \frac{dw}{2\pi i}\left(
\frac i2 \frac{(S^I \gamma^{ij} S_I)}{(\partial X^+)^2}\left(
\frac12\partial X^\mu \partial X_\mu
-\frac12 S_b^J\partial S_J^b
+
\frac74 
\frac{\partial^3 X^+}{\partial X^+}
-\frac34 
\frac{(\partial^2 X^+)^2}{(\partial X^+)^2} 
\right)\right.\nonumber\\
&&\hspace{8ex}\left.
+\frac34 i\frac{(S^I \gamma^{ij} \partial^2 S_I)}{(\partial X^+)^2}
\right),
\end{eqnarray}
which does not coincide with (\ref{Mi-Mj-6dGS}).

\section{The Lorentz invariance of the lower-dimensional DS superstrings}
\subsection{The $D=4$ DS superstring}
We now focus on the issue of the Lorentz invariance of 
the lower-dimensional DS superstrings
studied in Ref.~\citen{KM}.
We first briefly review the relevant results in the $D=4$ case.
The Lagrangian of the $D=4$ DS superstring is 
\begin{subequations}
\begin{eqnarray}
\mathcal{L}&=&\mathcal{L}_K+\mathcal{L}_{WZ},\\
\mathcal{L}_K&=&
-\frac{1}{2}\sqrt{-g}g^{ab}\Pi^\mu_a\Pi_{\mu b},\\
\mathcal{L}_{WZ}&=&
\epsilon^{ab}\Pi^\mu_a(W_{\mu b}-\hat{W}_{\mu b})
-\epsilon^{ab}W^\mu_a\hat{W}_{\mu b},
\end{eqnarray}
\end{subequations}
with
\begin{eqnarray}\label{Pi_a}
 \Pi^\mu_a&=&\partial_aX^\mu
-\sum_{A=1}^2i\partial_a(\theta^A\sigma^\mu\tilde{\bar{\theta}}^A
-\tilde{\theta}^A\sigma^\mu\bar{\theta}^A)-\sum_{A=1}^2W^{A\mu}_a
\end{eqnarray}
and
\begin{subequations}
\label{WA}
\begin{eqnarray}
 W^{A\mu}_a&=&i\Theta^A\sigma^\mu\partial_a\bar{\Theta}^A
-i\partial_a\Theta^A\sigma^\mu\bar{\Theta}^A,\\
\Theta^A&=&\tilde{\theta}^A-\theta^A,\qquad
\bar{\Theta}^A=\tilde{\bar{\theta}}^A-\bar{\theta}^A.
\end{eqnarray}
\end{subequations}
Here, $\tilde\theta^A$ and $\bar{\tilde\theta}^A$ are 
the spinors newly added to the GS superstring, and if 
they are set to zero, the Lagrangian reduces to that of the GS superstring.
Following Ref.~\citen{AK}, we impose the semi-light-cone gauge condition 
only on the spinors with tildes and compute the Dirac bracket. Then, 
we obtain a new set of canonical variables with respect to the Dirac bracket, 
in terms of which the remaining holomorphic first-class constraints read
as follows\cite{KM}:
\begin{subequations}\label{classicalconstraints}
\begin{align}
D_1=&d_1-i\sqrt{2\pi^+}\bar{S},\\
D_2=&d_2-i\sqrt{\frac{2}{\pi^+}}\pi\bar{S}
-\frac{2}{\pi^+}S\bar{S}\partial\bar{\theta}^{\dot 2},\\
\bar{D}_{\dot 1}=&\bar{d}_{\dot 1}+i\sqrt{2\pi^+}S,\\ 
\bar{D}_{\dot 2}=&\bar{d}_{\dot 2}+i\sqrt{\frac{2}{\pi^+}}\bar{\pi}S
+\frac{2}{\pi^+}S\bar{S}\partial\theta^2,\\
\mathcal{T}=&
-\frac{1}{2}\frac{\pi^\mu\pi_\mu}{\pi^+}
-\frac{1}{2}\frac{S\partial\bar{S}}{\pi^+}
+\frac{1}{2}\frac{\partial S\bar{S}}{\pi^+}
+i\sqrt{\frac{2}{\pi^+}}(S\partial\bar{\theta}^{\dot 1}
+\partial\theta^1\bar{S})
\nonumber\\
&
+i\sqrt{\frac{2}{(\pi^+)^3}}
\left(\bar{\pi}S\partial\bar{\theta}^{\dot2}
+\pi\partial\theta^2\bar{S}\right)
+4\frac{S\bar{S}\partial\theta^2\partial\bar{\theta}^{\dot2}}{(\pi^+)^2},
\end{align}
\end{subequations}
where 
\begin{subequations}\label{d4}
\begin{eqnarray}
d_\alpha&=&p_\alpha-i\partial X^\mu(\sigma_\mu\bar{\theta})_\alpha
-\frac{1}{2}\left(
(\theta\sigma^\mu\partial\bar{\theta})
-(\partial\theta\sigma^\mu\bar{\theta})\right)
(\sigma_\mu\bar{\theta})_\alpha,\\
\bar{d}_{\dot\alpha}&=&\bar{p}_{\dot\alpha}
-i\partial X^\mu(\theta\sigma_\mu)_{\dot\alpha}
-\frac{1}{2}\left(
(\theta\sigma^\mu\partial\bar{\theta})
-(\partial\theta\sigma^\mu\bar{\theta})\right)
(\theta\sigma_\mu)_{\dot\alpha},\\
\pi^\mu&=&i\partial X^\mu+\theta\sigma^\mu\partial\bar\theta-\partial\theta\sigma^\mu\bar\theta.
\end{eqnarray}
\end{subequations}
Here, the symbol $\partial$ represents $\frac\partial{\partial z}$. 
The quantities 
$\pi^\pm=\pi^0\pm\pi^3$, $\pi=\pi^1+i\pi^2$ and $\bar\pi=\pi^1-i\pi^2$ 
are also introduced.

The relevant OPEs among the basic fields are all free:
\begin{subequations}
\begin{align}
 X^\mu(z)X^\nu(w)\sim&\eta^{\mu\nu}\log(z-w),\\
p_\alpha(z)\theta^\beta(w)\sim&\frac{\delta_\alpha^\beta}{z-w},\\
\bar{p}_{\dot\alpha}(z)\bar{\theta}^{\dot\beta}(w)
\sim&\frac{\delta_{\dot\alpha}^{\dot\beta}}{z-w},\\
S(z)\bar{S}(w)\sim&\frac{1}{z-w}.
\end{align}
\end{subequations}
Again, the algebras of the constraints (\ref{classicalconstraints}) are not closed,
due to the presence of multiple contractions in the OPEs,
and this prevents us from constructing a nilpotent BRST charge. 
To remedy this, as in Ref.~\citen{AK}, we modify the constraints as 
\begin{subequations}\label{qc4}
\begin{alignat}{3}
&D_1 & \quad \rightarrow \quad
&\check D_1 & \quad \equiv \quad
&D_1,\\
&\bar{D}_{\dot1} & \quad \rightarrow \quad
&\check{\bar D}_{\dot1} & \quad \equiv \quad
&\bar{D}_{\dot1},\rule{0ex}{4ex}\\
&D_2 & \quad \rightarrow \quad
&\check D_2 & \quad \equiv \quad
&D_2-\frac{\partial^2\bar{\theta}^{\dot2}}{\pi^+}
+\frac{1}{2}\frac{\partial\pi^+\partial\bar{\theta}^{\dot2}}{(\pi^+)^2},\\
&\bar{D}_{\dot2} & \quad \rightarrow \quad
&\check{\bar D}_{\dot2} & \quad \equiv \quad
&\bar{D}_{\dot2}-\frac{\partial^2\theta^2}{\pi^+}
+\frac{1}{2}\frac{\partial\pi^+\partial\theta^2}{(\pi^+)^2},\\
&\mathcal{T} & \quad \rightarrow \quad
&\check{\mathcal{T}} & \quad \equiv \quad
&\mathcal{T}+
\frac{\partial\theta^2\partial^2\bar{\theta}^{\dot2}}{(\pi^+)^2}
-\frac{\partial^2\theta^2\partial\bar{\theta}^{\dot2}}{(\pi^+)^2}
-\frac{1}{8}\frac{\partial^2\log\pi^+}{\pi^+}.
\end{alignat}
\end{subequations}
The additional terms above can be viewed as arising from the normal-ordering 
ambiguities of  the constraints, and the precise values of the
coefficients have been determined so that the algebras are closed.
One can verify that these modified constraints have the OPE
\begin{equation}
\check{D}_2(z)\check{\bar{D}}_{\dot 2}(w) 
\sim \frac{4\check{\mathcal{T}}(w)}{z-w},
\end{equation}
without higher singularities, and is regular otherwise. 
In this way, we obtain a
set of first-class constraints which can be used to construct a nilpotent 
BRST charge in a conventional manner as
\begin{equation}
\tilde{Q}=\oint\frac{dz}{2\pi i}\left(
\lambda^\alpha\check{D}_\alpha
+\bar{\lambda}^{\dot\alpha}\check{\bar{D}}_{\dot\alpha}
+c\check{\mathcal{T}}
-4\lambda^2 \bar{\lambda}^{\dot2}b
\right).\label{BRSilde4}
\end{equation}
Here, $b$ and $c$ are the usual fermionic ghosts, satisfying 
\begin{eqnarray}
b(z)c(w)\sim\frac1{z-w},
\end{eqnarray}
while $\lambda^\alpha$ and 
$\bar{\lambda}^{\dot{\alpha}}$
are \textit{unconstrained} bosonic spinor ghosts, a part of which is 
identified as the pure spinor ghosts after the similarity transformations 
described in the next section.\footnote{Note that the $D=4$ pure spinor condition 
implies 
$ \lambda^\alpha=0$ or $\bar\lambda^{\dot\alpha}=0$, 
treating them as independent quantities (rather than complex conjugates),
as usual in the PS formalism. 
}

Let us now consider the Lorentz generators. All of them but $N^{i-}$ 
are obtained by adding generators constructed from 
$p$, $\theta$, $\lambda$ and $\omega$, the conjugate 
of $\lambda$, with
\begin{subequations}
\begin{eqnarray}
\lambda^\alpha(z)\omega_\beta(w)&\sim\frac{\delta^\alpha_\beta}{z-w},\\
\bar\lambda^{\dot\alpha}(z)\bar\omega_{\dot\beta}(w)&\sim\frac{\delta^{\dot\alpha}_{\dot\beta}}{z-w},
\end{eqnarray}
\end{subequations}
to those of the GS superstring in the semi-light-cone gauge,
\begin{subequations}
\begin{eqnarray}
N^{ij}&=&\frac14\left(-X^i\partial X^j +X^j\partial X^i +i\epsilon^{ij}S\bar S
\right.\nonumber\\
&&\hspace{8ex}\left.
+i\epsilon^{ij}(\theta\sigma_3 p
+\bar p\sigma_3 \bar\theta
-
\lambda\sigma_3 \omega
+
\bar \omega\sigma_3\bar\lambda)
\right),\\
N^{+-}&=&
\frac14\left(
-X^+\partial X^- +X^-\partial X^+ 
\right.\nonumber\\
&&\hspace{8ex}\left.
+2(\theta\sigma_3 p
-\bar p\sigma_3 \bar\theta
-
\lambda\sigma_3 \omega
-
\bar \omega\sigma_3\bar\lambda)
+4bc
\right),\\
N^{i+}&=&\frac14\left(
-X^i\partial X^+ +X^+\partial X^i
\right.\nonumber\\
&&\hspace{8ex}\left.
+2(s_i \theta^2 p_1 +\bar s_i \bar\theta^{\dot 2}\bar p_{\dot 1}
-
s_i \lambda^2 \omega_1 -\bar s_i \bar\lambda^{\dot 2}\bar \omega_{\dot 1}
) 
\right),
\end{eqnarray}
\end{subequations}
where 
\begin{eqnarray}
s_i=\left\{\begin{array}{ll}
1&(i=1)\\ i&(i=2)\\
\end{array}
\right.
~~\mbox{and}~~
\bar s_i=\left\{\begin{array}{ll}
1&(i=1)\\ -i&(i=2)\\
\end{array}
\right.. 
\end{eqnarray}
The generator $N^{+-}$ also contains a contribution from the $bc$-ghost.
This is because that these ghost fields are not Lorentz scalars,
which can be seen from the form of the BRST charge (\ref{BRSilde4}).
On the other hand, $N^{i-}$ involves extra terms coming from the compensating 
$\kappa$ symmetry, and also other terms for the BRST invariance. The result is
\begin{eqnarray}
N^{i-}&=&
\frac14\left(
-X^i\partial X^- + X^-\partial X^i 
+2 (\bar s_i \theta^1 p_2 + s_i \bar\theta^{\dot 1}\bar p_{\dot 2}
-\bar s_i \lambda^1 \omega_2 - s_i \bar\lambda^{\dot 1}\bar \omega_{\dot 2})
\rule{0ex}{4ex}\right.
\nonumber\\
&&\hspace{4ex}
+\frac{4\pi^i bc}{\pi^+}
+2i\epsilon^{ij}\frac{\pi^j S\bar S}{\pi^+}
+4\sqrt{2}i\frac{bc(\bar s_i S \partial\bar\theta^{\dot 2}+
s_i \partial \theta^2 \bar S)}{(\pi^+)^{\frac32}}
-\frac32\frac{\partial\pi^i}{\pi^+}\nonumber\\
&&\hspace{4ex}
-2\sqrt{2}i \frac{\bar s_i \partial S \partial \bar\theta^{\dot 2}
+s_i \partial\theta^2\partial\bar S}
{(\pi^+)^{\frac32}}
-\sqrt{2}i \frac{\bar s_i S \partial\bar\theta^{\dot 2}+
s_i \partial \theta^2 \bar S}{(\pi^+)^{\frac52}} \partial \pi^+
\nonumber\\
&&\hspace{4ex}
\left.
+12i\epsilon^{ij}\frac{\pi^j\partial\theta^2\partial\bar\theta^{\dot 2}}{(\pi^+)^2}
+6\frac{\bar s_i\partial\theta^1\partial\bar\theta^{\dot 2}-s_i\partial\theta^2\partial\bar\theta^{\dot 1}}
{\pi^+}
\right.\nonumber\\
&&\hspace{4ex}
\left.
+4\sqrt{2}i \frac{b(\bar s_i S \partial\bar\lambda^{\dot 2}-
s_i \partial \lambda^2 \bar S)}{(\pi^+)^{\frac12}}
\right). 
\end{eqnarray}
With the exception of $[M^{i-},M^{j-}]$,
these generators form the correct $D=4$ Lorentz algebra:
\begin{eqnarray}
{[}M^{\mu\nu},~M^{\rho\sigma}{]}
&=&-\frac12(\eta^{\nu\rho}M^{\mu\sigma}
-\eta^{\mu\rho}M^{\nu\sigma}
-\eta^{\nu\sigma}M^{\mu\rho}
+\eta^{\mu\sigma}M^{\nu\rho}
),\\
M^{\mu\nu}&\equiv&\oint\frac{dz}{2\pi i}N^{\mu\nu}(z).
\end{eqnarray}
The commutator ${[}M^{i-},~M^{j-}{]}$ is given by
\begin{eqnarray}
{[}M^{i-} ,~M^{j-}{]}&=&\oint\dz
\left[
\frac12
i\e \frac{\pp \pi^- - \pi\pb}{(\pp)^2} S\Sb
+
\frac34
\frac{-\pi^i\partial\pi^j +\pi^j\partial\pi^i}{(\pp)^2}
\right.
\nonumber\\
&&
+
\sqrt{2}\e bc\left(
\frac{\pb S\dthb^{\dot 2}-\pi\dth^2\Sb}{(\pp)^{5/2}}
+\frac{S\dthb^{\dot 1}-\dth^1 \Sb}{(\pp)^{3/2}}
\right)\nonumber\\
&&
-
 i \e \partial\left(\frac{bc}{\pp}\right) \frac{S\Sb}{\pp}
+
4 i \e\left(
-2\frac{\partial(bc)}{(\pp)^3}
+\frac{bc\partial \pp}{(\pp)^4}
\right)\dth^2\dthb^{\dot 2}
\nonumber\\
&&
+\sqrt{2}\e\left(
\frac32
\frac{\partial\pb S \dthb^{\dot 2}-\partial\pi\dth^2 \Sb}{(\pp)^{5/2}}
-
\frac12
\frac{\pb \partial S \dthb^{\dot 2}-\pi\dth^2 \partial\Sb}{(\pp)^{5/2}}
\right.\nonumber\\
&&\left.
~~~-
\frac74
\frac{(\pb  S \dthb^{\dot 2}-\pi\dth^2 \Sb)\partial\pp}{(\pp)^{7/2}}
-
\frac12
\frac{\partial S\dthb^{\dot 1}-\dth^1\partial\Sb}{(\pp)^{3/2}}
-
\frac14
\frac{(S\dthb^{\dot 1}-\dth^1\Sb)\partial\pp}{(\pp)^{5/2}}
\right)\nonumber\\
&&+
3
 i \e \left(
\frac{(\pp\pi^- -2\pi\pb)\dth^2\dthb^{\dot 2}}{(\pp)^2}
-\frac{\pb\dth^1\dthb^{\dot 2}+\pi\dth^2\dthb^{\dot 1}}{(\pp)^2}
-\frac{\dth^1\dthb^{\dot 1}}{\pp}
\right)\nonumber\\
&&
-
 i\e\left(
\frac{\partial^2\dth^2\partial^2\dthb^{\dot 2}}{(\pp)^3}
+\frac{\partial\pp\partial(\dth^2\dthb^{\dot 2})}{(\pp)^4}
-4\frac{(\partial\pp)^2\dth^2\dthb^{\dot 2}}{(\pp)^5}\right)
\nonumber\\
&&
+
\sqrt{2} \e\left(
\frac{b(\pb S\lb + \pi \ld \Sb)}{(\pp)^{3/2}}
+\frac{b(S\bar\lambda^{\dot 1}+\lambda^1 \Sb)}{\sqrt{\pp}}
\right)\nonumber\\
&&
+
2
i \e\left(
2\partial b\frac{\dthb^{\dot 2}\ld-\dth^2 \lb}{(\pp)^2} 
+ b\frac{\partial^2\bar\theta^{\dot 2}\ld-\partial^2 \theta^2 \lb}{(\pp)^2} 
- b\frac{\partial\bar\theta^{\dot 2}\ld-\partial \theta^2 \lb}{(\pp)^3}\partial\pp 
\right)\nonumber\\
&&+
2
i\e\frac{-\partial(S\dthb^{\dot 2})\dth^2 \Sb+\partial(\dth^2 \Sb)S\dthb^{\dot 2}}{(\pp)^3}
\nonumber\\
&&\left.
+
4
 i\e\frac{bS\Sb (\ld\dthb^{\dot 2}+\dth^2\lb)}{(\pp)^2}\right].
\label{Mi-Mj-4dDS}
\end{eqnarray}

We can show that the right hand side cannot be written 
in a BRST exact form as follows. First,
suppose that all the terms in (\ref{Mi-Mj-4dDS}) could be written in the form
\begin{eqnarray}
{[}\tilde{Q},~\oint\dz\mbox{(parent)}{]}
\end{eqnarray}
for some (parent). Then, note that (parent) cannot contain $\omega_\alpha$
or $\bar{\omega}_{\dot\alpha}$, 
because (\ref{Mi-Mj-4dDS}) contains neither  
$\omega_\alpha$ and $\bar{\omega}_{\dot\alpha}$ nor 
$p_\alpha$ and $\bar p_{\dot\alpha}$, which necessarily follows 
from the contraction with  $\lambda^\alpha d_\alpha
+\bar{\lambda}^{\dot\alpha} \bar{d}_{\dot\alpha}$. 
In analogy to the previous section, let us focus on terms that 
do not contain any of $\lambda$, $\bar\lambda$, $\partial \theta$ and 
$\partial\bar\theta$: 
\begin{eqnarray}
{[}M^{i-} ,~M^{j-}{]}&=&\oint\dz
\left(
\epsilon^{ij}\frac{iS\bar S}{\pi^+}
\left(
\frac12\frac{\pi^\mu \pi_\mu}{\pi^+}
-\partial\left(
\frac{bc}{\pi^+}
\right)
\right)
+\frac34
\frac{-\pi^i\partial \pi^j+\pi^j\partial \pi^i}{(\pi^+)^2}
\right) \nonumber\\
&&+{\cal O}(\partial\theta) +{\cal O}(\lambda).
\label{Mi-Mj-4dDSw/olambda}
\end{eqnarray}
This contribution could only arise from contraction with
$c\check{\mathcal{T}}$, and thus (parent) must contain $b$.
Taking into account the $\pi^+$ dependence of (\ref{Mi-Mj-4dDSw/olambda}), 
the $S\bar S$ terms can only arise from the OPE
between terms of $c{\mathcal{T}}$ ($\equiv c \mathcal{T}_0$),
that are independent of
both $\partial\theta$ and $\partial\bar\theta$,
and $\epsilon^{ij}\frac{S\bar S}{\pi^+}$. However, we find
\begin{eqnarray}
\left[\oint\dz c {\mathcal T}_0 ,~ \dw\left(
-i\epsilon^{ij}
\frac{bS\bar S}{\pi^+}
\right)
\right]&=&\oint\dz
\left(
\epsilon^{ij}\frac{iS\bar S}{\pi^+}
\left(
\frac12\frac{\pi^\mu \pi_\mu}{\pi^+}
-\partial\left(
\frac{bc}{\pi^+}
\right)
\right)
\right.
\nonumber\\
&&
~~~~~~~~~
-\frac38\frac{\partial^2 \pi^+}{(\pi^+)^2}
+\frac{15}8\frac{(\partial\pi^+)^2}{(\pi^+)^3}
\nonumber\\
&&
\left.~~~~~~~~~
-\frac34 i \frac{\partial^2 S\bar S + S\partial^2 \bar S}{(\pi^+)^2}
\right), 
\end{eqnarray}
which is inconsistent with (\ref{Mi-Mj-4dDSw/olambda}). Therefore,
the commutator $(\ref{Mi-Mj-4dDS})$ is not BRST-exact. 
Thus we have shown that the $D=4$ DS superstring has only 
partial Lorentz invariance, like the $D=4$ $GS$ superstring in the 
light-cone or semi-light-cone gauge.


\subsection{The $D=6$ DS superstring}

The Lagrangian of the $D=6$ DS superstrings is similarly given by
\begin{subequations} 
\begin{align}
\mathcal{L}_K=&
-\frac{1}{2}\sqrt{-g}g^{mn}\Pi^\mu_m\Pi_{\mu n},\\
\mathcal{L}_{WZ}=&
\epsilon^{mn}\Pi^\mu_m(W_{\mu n}-\hat{W}_{\mu n})
-\epsilon^{mn}W^\mu_m\hat{W}_{\mu n}, 
\end{align}
\end{subequations}
where
\begin{align}
 \Pi^\mu_m=&\partial_mX^\mu
-\sum_{A=1}^2i\partial_m(\theta^{IA}C\gamma^\mu\tilde{\theta}^A_I)
-\sum_{A=1}^2W^{A\mu}_m,\\
 W^{A\mu}_m=&i(\Theta^{IA}C\gamma^\mu\partial_m\Theta^A_I),\\
\Theta^A_I=&\tilde{\theta}^A_I-\theta^A_I.
\end{align}
Here we use the same convention as in Ref.~\citen{KM},
except that, for later convenience, we put a bar 
on the lower component in the light-cone 
decomposition of a $SU(2)$ Majorana-Weyl (MW) spinor:   
\begin{equation}
\theta_I^\alpha=
\begin{pmatrix}
\theta^a_I\\
\bar{\theta}^{\dot{a}}_I\\ 
\end{pmatrix},\qquad (a,\dot{a}=1,2)
\label{barnotation}
\end{equation}
where $a$ and $\dot{a}$ are the spinor indices of 
the transverse rotation $SO(4)\sim SU(2)\times SU(2)$.
The $SU(2)$ MW condition is given by
\begin{subequations}
\begin{align}
(\theta^a_I)^*=&\epsilon^{IJ}\theta^b_J\epsilon_{ba}\equiv \theta^I_a,\\
(\bar{\theta}^{\dot{a}}_I)^*=&\epsilon^{IJ}\bar{\theta}^{\dot{b}}_J\epsilon_{\dot{b}\dot{a}}
\equiv \bar{\theta}^I_{\dot{a}}. 
\end{align}
\end{subequations}

After some field redefinitions, we find that
the constraint generators are classically given by
\begin{subequations}
\begin{align}
D^I_a=&d^I_a+\sqrt{2\pi^+}S^I_a,\\
\bar{D}^I_{\dot{a}}=&\bar{d}^I_{\dot{a}}
+\sqrt{\frac{2}{\pi^+}}\pi^i(S^I\bar{\gamma}^i)_{\dot a}
+\frac{2}{\pi^+}S^I_bS^b_J\partial\bar{\theta}^J_{\dot{a}},
\\
\mathcal{T}=&
-\frac{1}{2}\frac{\pi^\mu\pi_\mu}{\pi^+}
-\frac{1}{2}\frac{S^J_a\partial S^a_J}{\pi^+}
-\sqrt{\frac{2}{\pi^+}}\partial\theta^J_aS^a_J
\nonumber\\
&\hspace{20mm}
-\sqrt{\frac{2}{\pi^+}}
\frac{\pi^i(\partial\bar{\theta}^J\gamma^iS_J)}{\pi^+}
+2\frac{\partial\bar{\theta}^I_{\dot a}\partial\bar{\theta}^{\dot a}_J
S^J_aS^a_I}{(\pi^+)^2},
\end{align}
\end{subequations}
where the super-covariant currents $d^I_\alpha$ and $\pi^\mu$
are defined by
\begin{subequations}\label{scc6}
\begin{align}
 d_\alpha^I=&p^I_\alpha+i\partial X^\mu(C\gamma_\mu\theta^I)_\alpha
+\frac{1}{2}(\theta^JC\gamma^\mu\partial\theta_J)
(C\gamma_\mu\theta^I)_\alpha,\label{d6}\\
\pi^\mu=&i\partial X^\mu+(\theta^IC\gamma^\mu\partial\theta_I).
\label{pi6}
\end{align}
\end{subequations}
The redefined fields are free and satisfy the relations
\begin{subequations}
\begin{align}
 X^\mu(z)X^\nu(w)\sim&\eta^{\mu\nu}\log(z-w),\\
p^I_\alpha(z)\theta^\beta_J(w)\sim&
\frac{\delta^I_J\delta_\alpha^\beta}{z-w},\\
S^a_I(z)S^b_J(w)\sim&-\frac{\epsilon_{IJ}\epsilon^{ab}}{z-w}.
\end{align}
\end{subequations}
Including quantum corrections,
we define $\check{D}^I_a$, $\check{\bar{D}}^I_{\dot{a}}$
and $\check{\mathcal{T}}$ as
\begin{subequations}
\begin{eqnarray}
\check{D}^I_a&=&D^I_a,\\
\check{\bar{D}}^I_{\dot{a}}&=&\bar{D}^I_{\dot{a}}
-2\frac{\partial^2\bar{\theta}^I_{\dot{a}}}{\pi^+}
+\frac{\partial\pi^+\partial\bar{\theta}^I_{\dot{a}}}{(\pi^+)^2}
+\frac{8}{3}\frac{\partial\bar{\theta}^I_{\dot{b}}\partial\bar{\theta}^{\dot{b}}_J
\partial\bar{\theta}^J_{\dot{a}}}{(\pi^+)^2}, \label{6dDadot}\\
\check{\mathcal{T}}&=&\mathcal{T}
-\frac{1}{4}\frac{\partial^2\log\pi^+}{\pi^+}
-2\frac{\partial^2\bar{\theta}^J_{\dot{b}}\partial\bar{\theta}^{\dot{b}}_J}
{(\pi^+)^2}
+\frac{8}{3}\frac{\partial\bar{\theta}^I_{\dot{a}}\partial\bar{\theta}^{\dot{a}}_J
\partial\bar{\theta}^J_{\dot{b}}\partial\bar{\theta}^{\dot{b}}_I}{(\pi^+)^3},\label{6dT}
\end{eqnarray}
\end{subequations}
then they satisfy
\begin{subequations}
\begin{align}
\check{\bar{D}}^I_{\dot{a}}(z)
\check{\bar{D}}^J_{\dot{b}}(w)\sim&
-\frac{4\epsilon^{IJ}\epsilon_{\dot{a}\dot{b}}
\check{\mathcal{T}}(w)}{z-w},\\
\textrm{[all other combinations]}\sim&0.
\end{align}
\end{subequations}

The BRST charge can be straightforwardly constructed from this
constraint algebra as
\begin{equation}
\tilde{Q}=\oint\dz\left(
\lambda^\alpha_I
\check{D}^I_\alpha
+c
\check{\mathcal{T}}
-2\bar{\lambda}^I_{\dot{a}}\bar{\lambda}^{\dot{a}}_Ib\right),
\label{BRSilde6}
\end{equation}
with the unconstrained bosonic ghost pair $\lambda^\alpha_I$
and $\omega^I_\alpha$ and the fermionic ghost pair $b$ and $c$, with 
\begin{subequations}
\begin{align}
c(z)b(w)\sim&\frac{1}{z-w},\\
\lambda^\alpha_I(z)\omega_\beta^J(w)
\sim&\frac{\delta^\alpha_\beta\delta^J_I}{z-w}. 
\end{align}
\end{subequations}
The BRST charge given in (\ref{BRSilde6}) is exactly nilpotent.

Using the light-cone decomposition, it is convenient to use
the rewritten forms
\begin{subequations}
\begin{eqnarray}
\pi^+&=&i\partial X^++2\bar{\theta}^I_{\dot a}
\partial\bar{\theta}^{\dot a}_I,\\
\pi^-&=&i\partial X^-+2\theta^I_a
\partial\theta^a_I,\\
\pi^i&=&i\partial X^i+(\partial\bar{\theta}^I\gamma^i\theta_I)
-(\bar{\theta}^I\gamma^i\partial\theta_I),\\
d^a_I&=&p^a_I-i\partial X^+\theta^a_I-i\partial X^i(\bar{\gamma}^i\theta^I)^a
+\partial\bar{\theta}^{\dot b}_J\bar{\theta}^J_{\dot b}\theta^a_I
-\bar{\theta}^{\dot b}_I\partial\bar{\theta}^J_{\dot b}\theta^a_J
+\bar{\theta}^{\dot b}_I\bar{\theta}^J_{\dot b}\partial\theta^a_J,\\
\bar{d}^{\dot a}_I&=&\bar{p}^{\dot a}_I
-i\partial X^-\bar{\theta}^{\dot a}_I
-i\partial X^i(\gamma^i\theta^I)^{\dot a}
+\partial\theta^b_J\theta^J_b\bar{\theta}^{\dot a}_I
-\theta^b_I\partial\theta^J_b\bar{\theta}^{\dot a}_J
+\theta^b_I\theta^J_b\partial\bar{\theta}^{\dot a}_J,
\end{eqnarray}
\end{subequations}
where 
we have used the notation
\begin{subequations} 
\begin{eqnarray}
\gamma_i=i\sigma_i~~~(i=1,2,3), &~~~&\gamma_4=1_2,\\
\bar{\gamma}_i=-i\sigma_i~~~(i=1,2,3), &~~~&\bar\gamma_4=1_2,
\end{eqnarray}
\end{subequations}
which are 2$\times$2 blocks of the gamma matrices defined in
Ref.~\citen{KM}.\footnote{These are denoted by $\tilde{\gamma}_i$ and
$\tilde{\bar{\gamma}}_i$ in Ref.~\citen{KM}.}
Their standard index positions are ${(\gamma_i)^{\dot a}}_b$ and 
${(\bar\gamma_i)^a}_{\dot b}$.
Using these 2$\times$2
matrices, we also define
\begin{subequations}
\begin{eqnarray}
{(\gamma_{ij})^a}_b&\equiv&-\frac i2
{(\bar\gamma_i\gamma_j -\bar\gamma_j \gamma_i)^a}_b,\\
{(\bar\gamma_{ij})^{\dot a}}_{\dot b}
&\equiv&-\frac i2
{(
\gamma_i\bar\gamma_j 
-\gamma_j \bar\gamma_i
)^{\dot a}}_{\dot b},\\
{(\gamma_{ijk})^{\dot a}}_b&\equiv&+\frac16
{(
\gamma_i\bar\gamma_j\gamma_k -\gamma_i \bar\gamma_k\gamma_j
+\gamma_j\bar\gamma_k\gamma_i -\gamma_j \bar\gamma_i\gamma_k
+\gamma_k\bar\gamma_i\gamma_j -\gamma_k \bar\gamma_j\gamma_i
)
^{\dot a}}_{b}.
\end{eqnarray}
\end{subequations}

\begin{subequations}
The Lorentz generators, except for $N^{i-}$, can be easily obtained as
\begin{align}
N^{ij}=&
-\frac{1}{4}X^i\partial X^j+\frac{1}{4}X^j\partial X^i
+\frac{i}{4}(\theta^I\gamma^{ij}p_I)
+\frac{i}{4}(\bar{\theta}^I\bar{\gamma}^{ij}\bar{p}_I)
\nonumber\\
&\hspace{25mm}
-\frac{i}{4}(\lambda^I\gamma^{ij}\omega_I)
-\frac{i}{4}(\bar{\lambda}^I\bar{\gamma}^{ij}\bar{\omega}_I)
+\frac{i}{8}(S^I\gamma^{ij}S_I), \label{6dDSNij}
\\
N^{+-}=&
-\frac{1}{4}X^+\partial X^-
+\frac{1}{4}X^-\partial X^+
+\frac{1}{2}\theta^I_ap^a_I
-\frac{1}{2}\bar{\theta}^I_{\dot{a}}\bar{p}^{\dot{a}}_I
-\frac{1}{2}\lambda^I_a\omega^a_I
+\frac{1}{2}\bar{\lambda}^I_{\dot{a}}\bar{\omega}^{\dot{a}}_I
+bc,\label{6dDSN+-}
\\
N^{i+}=&
-\frac{1}{4}X^i\partial X^++\frac{1}{4}X^+\partial X^i
+\frac{1}{2}(\bar{\theta}^I\gamma^ip_I)
-\frac{1}{2}(\bar{\lambda}^I\gamma^i\omega_I).
\label{6dDSNi+}
\end{align}
The remaining generator $N^{i-}$ is given by
\begin{align}
N^{i-}=&
-\frac{1}{4}X^i\partial X^-+\frac{1}{4}X^-\partial X^i
+\frac{1}{2}(\theta^I\bar{\gamma}^i\bar{p}_I)
-\frac{1}{2}(\lambda^I\bar{\gamma}^i\bar{\omega}_I)
\nonumber\\
&\hspace{.5cm}
+\frac{\pi^ibc}{\pi^+}
+\frac{1}{4}\frac{i\pi^j(S^I\gamma^{ij}S_I)}{\pi^+}
-\frac{1}{4}\frac{\partial\pi^i}{\pi^+}
-\sqrt{2}\frac{bc(\partial\bar{\theta}^I\gamma^iS_I)}{(\pi^+)^{3/2}}
\nonumber\\
&\hspace{.5cm}
-\frac{\sqrt{2}}{3}\frac{(\partial\bar{\theta}^I\gamma^iS_J)S^J_aS^a_I}
{(\pi^+)^{3/2}}
+\frac{1}{\sqrt{2}}\frac{\partial\pi^+(\partial\bar{\theta}^I\gamma^iS_I)}
{(\pi^+)^{5/2}}
-\frac{i\pi^j(\partial\bar{\theta}^I\bar{\gamma}^{ij}\partial\bar{\theta}_I)}{(\pi^+)^2}
\nonumber\\
&\hspace{.5cm}
-\frac{(\partial\bar{\theta}^I\gamma^i\partial\theta_I)}{\pi^+}
-\frac{8\sqrt{2}}{3}\frac{(\partial\bar{\theta}^I\gamma^iS_J)\partial\bar{\theta}^J_{\dot{a}}
\partial\bar{\theta}^{\dot{a}}_I}{(\pi^+)^{5/2}}
+\sqrt{2}\frac{b(\bar{\lambda}^I\gamma^iS_I)}{(\pi^+)^{1/2}}.
\label{6dDSNi-}
\end{align}
\end{subequations}

The integrated generators
\begin{equation}
M^{\mu\nu}=\oint\dz N^{\mu\nu}(z)
\end{equation}
are BRST invariant, satisfying $[\tilde{Q},M^{\mu\nu}]=0$,
and form the Lorentz algebra, except for
\begin{align}
&[M^{i-},M^{j-}]
\nonumber\\
&=
\oint\dz\Bigg(
-\frac{1}{2}\left(
\delta^{ik}\delta^{jl}-\frac{1}{2}\epsilon^{ijkl}\right)
\frac{\pi^k\partial\pi^l-\pi^l\partial\pi^k}{(\pi^+)^2}
\nonumber\\
&\hspace{1.5cm}
+\frac{i}{2}\frac{(S^I\gamma^{ij}S_I)}{\pi^+}
\left(\frac{1}{2}\frac{\pi^\mu\pi_\mu}{\pi^+}
+\frac{1}{8}\frac{S^J_a\partial S^a_J}{\pi^+}
+\frac{1}{4}\frac{\partial^2\pi^+}{(\pi^+)^2}
-\partial\left(\frac{bc}{\pi^+}\right)
\right)
\nonumber\\
&\hspace{1.5cm}
-\frac{i}{4}\frac{(S^I\gamma^{ij}\partial^2 S_I)}{(\pi^+)^2}
-\frac{i}{8}\frac{(S^I\gamma^{ij}\partial S_J)S^J_aS^a_I}{(\pi^+)^2}
\nonumber\\
&\hspace{1.5cm}
+\sqrt{2}b\left(
\frac{\pi^i(\bar{\lambda}^I\gamma^j S_I)}{(\pi^+)^{3/2}}
-\frac{\pi^j(\bar{\lambda}^I\gamma^i S_I)}{(\pi^+)^{3/2}}
+\frac{\pi^k(\bar{\lambda}^I\gamma^{ijk}S_I)}{(\pi^+)^{3/2}}
+\frac{i(\lambda^I\gamma^{ij}S_I)}{(\pi^+)^{1/2}}
\right)
\nonumber\\
&\hspace{1.5cm}
-\sqrt{2}bc\left(
\frac{\pi^i(\partial\bar{\theta}^I\gamma^j S_I)}{(\pi^+)^{5/2}}
-\frac{\pi^j(\partial\bar{\theta}^I\gamma^i S_I)}{(\pi^+)^{5/2}}
+\frac{\pi^k(\partial\bar{\theta}^I\gamma^{ijk}S_I)}
{(\pi^+)^{5/2}}
+\frac{i(\partial\theta^I\gamma^{ij}S_I)}{(\pi^+)^{3/2}}
\right)
\nonumber\\
&\hspace{1.5cm}
-\frac{1}{\sqrt{2}}\Bigg(
3\frac{\partial\pi^i(\partial\bar{\theta}^I\gamma^j S_I)}
{(\pi^+)^{5/2}}
-3\frac{\partial\pi^j(\partial\bar{\theta}^I\gamma^i S_I)}
{(\pi^+)^{5/2}}
-4\frac{\partial\pi^+\pi^i(\partial\bar{\theta}\gamma^j S_I)}
{(\pi^+)^{7/2}}
\nonumber\\
&\hspace{3.5cm}
+4\frac{\partial\pi^+\pi^j(\partial\bar{\theta}\gamma^i S_I)}
{(\pi^+)^{7/2}}
-\frac{\partial\pi^+\pi^k(\partial\bar{\theta}^I\gamma^{ijk}S_I)}
{(\pi^+)^{7/2}}
-\frac{i\partial\pi^+(\partial\theta^I\gamma^{ij}S_I)}{(\pi^+)^{5/2}}
\Bigg)
\nonumber\\
&\hspace{1.5cm}
-2b\frac{i\partial\pi^+(\bar{\lambda}^I\bar{\gamma}^{ij}
\partial\bar{\theta}_I)}{(\pi^+)^3}
+\frac{b}{3}\Bigg(
8\frac{(\bar{\lambda}^I\gamma^i S_I)
(\partial\bar{\theta}^J\gamma^j S_J)}
{(\pi^+)^2}
-8\frac{(\bar{\lambda}^I\gamma^j S_I)
(\partial\bar{\theta}^J\gamma^i S_J)}
{(\pi^+)^2}
\nonumber\\
&\hspace{1.5cm}
-4\frac{(\bar{\lambda}^I\gamma^i S_J)
(\partial\bar{\theta}^J\gamma^j S_I)}
{(\pi^+)^2}
+4\frac{(\bar{\lambda}^I\gamma^j S_J)
(\partial\bar{\theta}^J\gamma^i S_I)}
{(\pi^+)^2}
-12\frac{i(\partial\bar{\lambda}^I\bar{\gamma}^{ij}
\partial\bar{\theta}_I)}{(\pi^+)^2}
+6\frac{i\partial\pi^+(\bar{\lambda}^I\bar{\gamma}^{ij}
\partial\bar{\theta}_I)}{(\pi^+)^3}
\nonumber\\
&\hspace{1.5cm}
-4\frac{i(\bar{\lambda}^I\bar{\gamma}^{ij}\partial\bar{\theta}_J)
S^J_aS^a_I}{(\pi^+)^2}
\Bigg)
-\frac{\sqrt{2}}{3}\Bigg(
\frac{\pi^i(\partial\bar{\theta}^I\gamma^j S_J)S^J_aS^a_I}
{(\pi^+)^{5/2}}
-\frac{\pi^j(\partial\bar{\theta}^I\gamma^i S_J)S^J_aS^a_I}
{(\pi^+)^{5/2}}
\nonumber\\
&\hspace{8cm}
+\frac{\pi^k(\partial\bar{\theta}^I\gamma^{ijk}S_J)S^J_aS^a_I}
{(\pi^+)^{5/2}}
+\frac{i(\partial\theta^I\gamma^{ij}S_J)S^J_aS^a_I}
{(\pi^+)^{3/2}}\Bigg)
\nonumber\\
&\hspace{1.5cm}
-\frac{i(\partial\bar{\theta}^I\bar{\gamma}^{ij}\partial\bar{\theta}_I)}
{(\pi^+)^2}\left(
\pi^-+\frac{S^J_b\partial S^b_J}{\pi^+}
+\frac{1}{2}\frac{\partial^2\log\pi^+}{\pi^+}
-4\frac{b\partial c}{\pi^+}\right)
\nonumber\\
&\hspace{1.5cm}
+2\frac{i\pi^i\pi^k(\partial\bar{\theta}^I\bar{\gamma}^{kj}
\partial\bar{\theta}_I)}{(\pi^+)^3}
+2\frac{i\pi^k\pi^j(\partial\bar{\theta}^I\bar{\gamma}^{ik}
\partial\bar{\theta}_I)}{(\pi^+)^3}
-2\frac{\pi^i(\partial\bar{\theta}^I\gamma^j\partial\theta_I)}
{(\pi^+)^2}
+2\frac{\pi^j(\partial\bar{\theta}^I\gamma^i\partial\theta_I)}
{(\pi^+)^2}
\nonumber\\
&\hspace{1.5cm}
-2\frac{\pi^k(\partial\bar{\theta}^I\gamma^{ijk}\partial\theta_I)}
{(\pi^+)^2}
-\frac{i(\partial\theta^I\gamma^{ij}\partial\theta_I)}{\pi^+}
-\frac{(\partial\bar{\theta}^I\gamma^i S_I)
\partial(\partial\bar{\theta}^J\gamma^j S_J)}
{(\pi^+)^3}
+\frac{(\partial\bar{\theta}^I\gamma^j S_I)
\partial(\partial\bar{\theta}^J\gamma^i S_J)}
{(\pi^+)^3}
\nonumber\\
&\hspace{1.5cm}
-\frac{(\partial\bar{\theta}^I\gamma^i S_J)
\partial(\partial\bar{\theta}^J\gamma^j S_I)}
{(\pi^+)^3}
+\frac{(\partial\bar{\theta}^I\gamma^j S_J)
\partial(\partial\bar{\theta}^J\gamma^i S_I)}
{(\pi^+)^3}
\nonumber\\
&\hspace{1.5cm}
+2\frac{i(\partial\bar{\theta}^I\bar{\gamma}^{ij}\partial^2\bar{\theta}_J)
S^J_aS^a_I}{(\pi^+)^3}
-\frac{32b}{3}\frac{i(\bar{\lambda}^I\bar{\gamma}^{ij}\partial\bar{\theta}_J)
\partial\bar{\theta}^J_{\dot a}\partial\bar{\theta}^{\dot
 a}_I}{(\pi^+)^3}
+\frac{i}{3}\frac{(S^I\gamma^{ij}S_I)S^J_aS^a_K
\partial\bar{\theta}^K_{\dot a}\partial\bar{\theta}^{\dot a}_J}
{(\pi^+)^3}
\nonumber\\
&\hspace{1.5cm}
-\frac{8\sqrt{2}}{3}\Bigg(
2\frac{\pi^i(\partial\bar{\theta}^I\gamma^jS_J)
\partial\bar{\theta}^J_{\dot a}\partial\bar{\theta}^{\dot a}_I}
{(\pi^+)^{7/2}}
-2\frac{\pi^j(\partial\bar{\theta}^I\gamma^iS_J)
\partial\bar{\theta}^J_{\dot a}\partial\bar{\theta}^{\dot a}_I}
{(\pi^+)^{7/2}}
+\frac{\pi^k(\partial\bar{\theta}^I\gamma^{ijk}S_J)
\partial\bar{\theta}^J_{\dot a}\partial\bar{\theta}^{\dot a}_I}
{(\pi^+)^{7/2}}\Bigg)
\nonumber\\
&\hspace{1.5cm}
+2\sqrt{2}\Bigg(
\frac{(\partial\bar{\theta}^I\gamma^i S_J)
(\partial\bar{\theta}_I\gamma^j\partial\theta^J)}
{(\pi^+)^{5/2}}
-\frac{(\partial\bar{\theta}^I\gamma^j S_J)
(\partial\bar{\theta}_I\gamma^i\partial\theta^J)}
{(\pi^+)^{5/2}}
-2\frac{i(\partial\theta^I\gamma^{ij}S_J)
\partial\bar{\theta}^J_{\dot a}\partial\bar{\theta}^{\dot a}_I}
{(\pi^+)^{5/2}}\Bigg)
\nonumber\\
&\hspace{1.5cm}
+\frac{i(\partial\bar{\theta}^I\bar{\gamma}^{ij}\partial\bar{\theta}_I)
\partial\bar{\theta}^J_{\dot a}\partial^2\bar{\theta}^{\dot a}_J}
{(\pi^+)^4}
-2\frac{i(\partial\bar{\theta}^I\bar{\gamma}^{ij}\partial^2\bar{\theta}_J)
\partial\bar{\theta}^J_{\dot a}\partial\bar{\theta}^{\dot a}_I}
{(\pi^+)^4}
+\frac{8}{3}\frac{i(\partial\bar{\theta}^I\bar{\gamma}^{ij}
\partial\bar{\theta}_I)\partial\bar{\theta}^J_{\dot a}
\partial\bar{\theta}^{\dot a}_KS^K_a S^a_J}{(\pi^+)^4}
\Bigg).
\end{align}
In particular, we have
\begin{align}
 &[M^{i-},M^{j-}]
\nonumber\\
&=
\oint\dz\Bigg(
-\frac{1}{2}\left(
\delta^{ik}\delta^{jl}-\frac{1}{2}\epsilon^{ijkl}\right)
\frac{\pi^k\partial\pi^l-\pi^l\partial\pi^k}{(\pi^+)^2}
\nonumber\\
&\hspace{1.5cm}
+\frac{i}{2}\frac{(S^I\gamma^{ij}S_I)}{\pi^+}
\left(\frac{1}{2}\frac{\pi^\mu\pi_\mu}{\pi^+}
+\frac{1}{8}\frac{S^J_a\partial S^a_J}{\pi^+}
+\frac{1}{4}\frac{\partial^2\pi^+}{(\pi^+)^2}
-\partial\left(\frac{bc}{\pi^+}\right)
\right)
\nonumber\\
&\hspace{1.5cm}
-\frac{i}{4}\frac{(S^I\gamma^{ij}\partial^2 S_I)}{(\pi^+)^2}
-\frac{i}{8}\frac{(S^I\gamma^{ij}\partial S_J)S^J_aS^a_I}{(\pi^+)^2}
\Bigg)+\mathcal{O}(\partial\theta)+\mathcal{O}(\lambda).
\end{align}
On the other hand, we find
\begin{align}
&\{\oint\dz c\mathcal{T}_0,\oint\dz\left(
-\frac{i}{2}\frac{b(S^I\gamma^{ij}S_I)}{\pi^+}\right)\}
\nonumber\\
&\hspace{1cm}
=\oint\dz\Bigg(
\frac{i}{2}\frac{(S^I\gamma^{ij}S_I)}{\pi^+}
\Bigg(
\frac{1}{2}\frac{\pi^\mu\pi_\mu}{\pi^+}
+\frac{1}{2}\frac{S^J_a\partial S^a_J}{\pi^+}
-\frac{1}{4}\frac{\partial^2\pi^+}{(\pi^+)^2}
+\frac{7}{4}\frac{(\partial\pi^+)^2}{(\pi^+)^3}
-\partial\left(\frac{bc}{\pi^+}\right)
\Bigg)
\nonumber\\
&\hspace{2.5cm}
-\frac{3}{4}\frac{i(S^I\gamma^{ij}\partial^2S_I)}{(\pi^+)^2}
\Bigg),
\end{align}
where, as above, $\mathcal{T}_0$ is the $\partial\theta$-independent part of 
${\mathcal{T}}$. Then, repeating the same argument as in the previous section, 
we find that the commutator is not BRST exact.

\section{Conclusions and discussion}
In this paper, we have shown that the $D=4$ and 6 double-spinor (DS) superstrings 
do not possess the full  Lorentz symmetry, as in the light-cone and semi-light-cone 
gauge quantizations of lower-dimensional Green-Schwarz superstrings. 

We have emphasized that the modification of the energy-momentum tensor 
is a common procedure employed to preserve quantum conformal invariance  
in the semi-light-cone gauge quantization, even in the critical case. 
One can rewrite the logarithmic term of the energy-momentum tensor
(\ref{Tmodified}) or, more generally, (\ref{TX+X-}) in
the usual linear-dilaton form by bosonization. Owing to the relation
\begin{eqnarray}
\partial X^+(z)X^-(w)\sim\frac 2 {z-w},
\end{eqnarray}
we can identify them as a $\beta\gamma$-system. Therefore, we define 
\begin{subequations} 
\begin{eqnarray}
\partial X^+(z)&=&\gamma(z)~=~e^{\phi-\chi}(x),\\
X^-(z)&=&2\beta(z)~=~2\partial \chi e^{-\phi-\chi}(x),
\end{eqnarray}
where $\gamma(z)\beta(w)\sim\frac1{z-w}$, 
$\phi(z)\phi(w)\sim -\log(z-w)$ and 
$\chi(z)\chi(w)\sim +\log(z-w)$.
\end{subequations} Plugging these into (\ref{TX+X-}), we obtain
\begin{eqnarray}
T_{X^+X^-}(z)&=&-\frac12 (\partial\phi)^2 +\left(
\frac12 +\xi
\right)\partial^2 \phi
+\frac12 (\partial\chi)^2 +\left(
\frac12 -\xi
\right)\partial^2 \chi,
\end{eqnarray}
where $\xi=\frac78$ ($D=4$), $\frac34$ ($D=6$) and $\frac12$ ($D=10$).
Therefore, the modification of the energy-momentum tensor can be regarded 
as a change of the background from flat to linear-dilaton, although the dilaton is 
only linear with respect to the special bosonized coordinates. This way of viewing 
the modification is consistent with that in recent works on the relation between the 
lower-dimensional PS and non-critical superstrings \cite{AGMOY}. 
It is also interesting that  the $\chi$ field becomes 
a normal scalar in the critical ($D=10$) case. 
However, the meaning of this observation 
is yet unclear.

We showed in Ref.~\citen{KM} that 
the physical spectra of the $D=4$ and $D=6$ DS superstrings 
coincide with those of the pure-spinor (PS) formalisms in the same 
numbers of dimensions.  
Let us now compare the Lorentz generators given in 
Refs.~\citen{GW} and \citen{W} and ours obtained in the DS formalism. 
In four dimensions, the necessary similarity transformations relating the BRST 
charges of the two $D=4$ theories are \cite{KM}
%
%
\begin{eqnarray}
X&=&-\frac{1}{4}\oint\frac{dz}{2\pi i}\frac{c\check{\bar{D}}_{\dot2}}
{\tilde{\lambda}^2},
\\
Y&=&-\frac{1}{2}\oint\frac{dz}{2\pi i}S\bar{S}\log\pi^+,\\
Z&=&\oint\frac{dz}{2\pi i}\left(
\frac{i}{\sqrt{2}}\bar{d}_{\dot1}\bar{S}
+\frac{\partial\theta^2\partial\bar{\theta}^{\dot2}}{\pi^+}\right).
\end{eqnarray}
Then, the BRST charge $\tilde{Q}$ is transformed to
\begin{eqnarray}
&&(e^Z e^Y e^X) \tilde{Q}(e^{Z} e^{Y} e^{X})^{-1}=Q+\delta_b+\delta,\\
&&Q=\oint\frac{dz}{2\pi i}\lambda^\alpha d_\alpha,\\
&&\delta_b=-4\oint\frac{dz}{2\pi i}{\lambda}^2{\bar{\lambda}}
^{\dot2}b,
\\
&&\delta=\sqrt{2}i\oint\frac{dz}{2\pi i}\bar{\lambda}^{\dot1}S,\qquad
\end{eqnarray}
where $\delta_b$ and $\delta$ anti-commute with $Q$ and have trivial 
cohomologies of the BRST quartets 
$(b,c;{\bar{\lambda}}^{\dot 2},{\bar{\omega}}_{\dot 2})$ and
$(S,\bar S;{\bar{\lambda}}^{\dot 1},{\bar{\omega}}_{\dot 1})$ 
(where $\bar \omega_{\dot\alpha}$ is the field conjugate
to $\bar\lambda^{\dot\alpha}$). One can 
alternatively decouple $\lambda^\alpha$ instead of $\bar\lambda^{\dot\alpha}$.
Taking the quotients with respect to 
the Hilbert space of these BRST trivial fields leaves 
precisely the $D=4$ PS  Hilbert space with the BRST charge $Q$ proposed 
in Refs.~\citen{GW} and \citen{W}.

The $D=4$ PS superstring has an anomaly-free set of level-1 Lorentz currents.
If they are similarity-transformed back to the DS theory by using the above $X$, 
$Y$ and $Z$, they do not coincide with the Lorentz generators we considered 
in the previous section. This is obvious, because the Lorentz generators in the PS 
formalism do not act on the BRST-quartet fields decoupled through the similarity 
transformations. This can also be verified by an explicit calculation. 
Thus, we conclude that, although the generators of the PS formalism realize a 
representation of the $D=4$ Lorentz group on the PS fields, they are not directly 
related to the symmetries of the DS Lagrangian.
A similar statement holds in the $D=6$ case.

\section*{Acknowledgements}
The work of HK is supported in part by a Grant-in-Aid
for Scientific Research (No.~19540284) and a Grant-in-Aid
for the 21st Century COE \lq\lq Center for Diversity and
Universality in Physics'', while  the work of SM is supported by 
a Grant-in-Aid for Scientific Research (No.16540273)
from the Ministry of Education,
Culture, Sports, Science and Technology (MEXT) of Japan.

\end{document}